\documentclass[a4paper,fleqn]{cas-dc}
\usepackage[numbers, compress]{natbib}

\def\tsc#1{\csdef{#1}{\textsc{\lowercase{#1}}\xspace}}
\tsc{WGM}
\tsc{QE}
\tsc{EP}
\tsc{PMS}
\tsc{BEC}
\tsc{DE}

\begin{document}
\let\WriteBookmarks\relax
\def\floatpagepagefraction{1}
\def\textpagefraction{.001}
\shorttitle{Scrambling of Entanglement from Integrability to Chaos}
\shortauthors{M S\"uzen}

\title [mode = title]{Scrambling of Entanglement from Integrability to Chaos:  \\
       Bootstrapped Time-Integrated Spread Complexity}                      


\author[1,2]{M. S\"uzen}[orcid=0000-0002-9460-7297]
\ead{mehmet.suzen@physics.org}
\affiliation[1]{organization={Resident Scientist},
                city={Assia},
                postcode={CY 5561}, 
                country={Cyprus}}
\affiliation[2]{organization={American Physical Society},
              addressline={Member}, 
                city={College Park},
                state={MD},
                country={United States}}


\credit{Conceptualization, Methodology, Software, Data Analysis, Visualisation, Writing}

\begin{abstract} 
A physical relationship for the combination of scrambling via spread
complexity and entanglement is characterized by the fidelity of quantum
unitary dynamics. Newly introduced time-integrated quantities for the
degree of quantum ergodicity captures state spreading from initial to
late times under physically plausible operator perturbations. Using
bootstrapped Hamiltonian realizations as a perturbation scheme shown
to be a powerful testbed. Selecting maximally entangled state as a
starting state captures the scrambling of entanglement in this setting.
For this reason, we utilize the Rosenzweig-Porter ensembles across
different ergodic regimes. Computed integrated spread complexity and
associated integrated quantum state fidelity displays a monotonic
inverse relationship for maximally entangled states from integrability
to chaos.
\end{abstract}


\begin{highlights}
\item Time-integrated spread complexity and state fidelity are linked via Krylov bases.
\item Scrambling is evaluated via time-integrated fidelity of a maximum entangled 
state.
\item A monotonic inverse relationship is formally derived and verified numerically.
\end{highlights}


\maketitle

\section{Introduction}

Probably the most fascinating ideas and practical impacts of quantum 
mechanics manifest in the interplay among entanglement 
\cite{einstein35, bell04, horo09, susskind15qm, ho17}, quantum state and operator 
complexity \cite{nielsen10}, and information scrambling to 
reach ergodicity \cite{hayden07, susskind08}. In this context, 
information scrambling is deeply intertwined with ergodicity 
and thermalization \cite{jdeutsch91, srednicki94, srednicki96}. 
Quantum chaos \cite{berry89, nakamura94, zurek95, gutzwiller13, 
potters20} provides a unifying theme for this landscape: 
from the perspective of random matrix theory, it offers a 
spectral analysis of quantum systems whose classical counterparts 
are chaotic in the Lyapunov sense \cite{wigner51, dyson62a, 
berry77, bohigas84, berry84, maldacena16, cotler17}.

Recent advances in measuring the complexity of quantum systems 
have addressed the operator growth hypothesis \cite{parker19} 
and Krylov complexity \cite{rabi22}, as well as quantum state
complexity quantified by the spread of dynamics \cite{bala22}. 
These frameworks have been applied across diverse many-body 
settings, including stadium billiards \cite{hashimoto}, Heisenberg-Ising 
chains \cite{camargo24}, density matrices \cite{chaputa24q}, 
quantum kicked rotor (QKR) \cite{kannan}, Gaussian random matrices \cite{erdmenger}, 
Rosenzweig-Porter random matrices \cite{bhat25}, the SYK (Sachdev-Ye-Kitaev) 
model \cite{bagg25}, basic qubit dynamics \cite{seet25}
and its comparison to Nielsen complexity \cite{craps24}. 
Building on these developments, and following the foundational 
ideas of Susskind regarding time-integrated circuit 
complexity \cite{susskind20}, we propose a global extension 
of spread complexity defined by its time integral. This 
measure provides a robust tool to distinguish between distinct 
ergodic regimes.

We investigate maximally entangled states evolving under unitary
dynamics across the spectrum from quantum chaos to integrability, 
monitoring both time-integrated spread complexity and fidelity. 
Our analysis reveals that time-integrated spread complexity acts 
as a sensitive probe for differentiating ergodic regimes. 
This approach enables a fine-grained distinction of quantum 
ergodicity, contributing to the fields of quantum chaos 
diagnosis \cite{santos04, santos12, das25} and 
information scrambling \cite{lomonaco25}.

In this direction, out-of-time-ordered correlators (OTOCs) 
\cite{larkin, swingle16ms, wolynes, sahu24, lomonaco25} 
and their echoes serve as standard measures of information 
scrambling \cite{iyoda18sq, lewis19, zurek20, mi21, googleq25, 
abd25, shekhar25a}. Here, we shift focus to the relationship between 
fidelity and spread complexity during the unitary evolution 
of maximally entangled qubits across varying degrees of 
ergodicity. Within our framework, we introduce numerically 
robust, bootstrapped integrated values by perturbing the 
Hamiltonian slightly to generate an ensemble of unitary 
evolutions. This approach serves two purposes: it 
provides a rigorous estimation of statistical uncertainty 
and assesses the robustness of state growth against
small dynamical perturbations due to different unitary 
evolutions imposed by the slightly different operator.

Probing the entanglement with quantum state
fidelity out-of-time-order correlators (FOTOCs) is recently introduced
\cite{lewis19a}. FOTOCs provides a composite view of
scrambling and entanglement. Due to limitations of out-of-time order
correlators on the choice of operators, spread complexity is more
objective measure in general \cite{rabi25, nandy25pr, baiguera26}. Moreover,
using time-integrated quantities would cover the early to late-time dynamics.
From these perspective, our approach provides a conceptual advance in 
combining scrambling and entanglement via leveraging the basic idea 
of spreading \cite{parker19, bala22}, introducing forward spread complexity 
and fidelity separation. Quantifying quantum ergodicity diagnosis 
via time-integrated fashion from early to late times gives a 
complementary measure capturing spreading over unitary evolution.

We probe entanglement scrambling by using a maximally entangled
state as the initial state for the unitary evolution. This will 
allow us representing a regime of full quantum correlation.
We compute the time-integrated spread complexity and
track it simultaneously with time-integrated fidelity,
leveraging their analytical correlation. This approach
captures scrambling insights comparable to FOTOCs.
Initializing the system in this state of full quantum
correlation reveals the inverse correlation between spread
complexity and fidelity much more clearly as the system
evolves, avoiding any bias associated with localized
initial product states.

Advancing the physical insights and mechanisms of non-equilibrium many-body
dynamics within the overlapping theme of quantum information theory
and high-energy physics are of fundamental interest. For this reason,
having a robust and easily accessible diagnostic for quantum ergodicity
and entanglement for information scrambling in complex quantum systems
is quite significant.

In Section \ref{sec:scram}, we provide the dynamical 
setting where scrambling of entanglement can be studied for 
unitary time evolution. In Section \ref{sec:spread},
details of spread complexity via Krylov bases is introduced.
We also provide an analytical prediction for connecting fidelity 
of quantum dynamics to spread complexity using Krylov bases
initial entangled state. We provide how Rosenzweig-Porter ensemble 
can be used as a perturbation scheme with bootstrapping the Hamiltonian
in Section \ref{sec:rp}, and our numerical results. Discussion 
of results are given in Section \ref{sec:dis} with broader implications. 
We conclude our work in Section \ref{sec:con}.

\section{Scrambling of entanglement: Dynamical Setting} \label{sec:scram}

There is an inherent algorithmic connection between randomness 
and complexity. Algorithmic information theory pioneered by 
Solomonoff, Kolmogorov, and Chaitin \cite{solomonoff64, 
kolmogorov65, chaitin66} (SKC) addresses quantification of 
this, so-called Kolmogorov complexity classically. 
In quantifying the complexity of quantum mechanical system 
different approaches are taken, such as the out-of-time-order 
scrambling \cite{susskind08, susskind20}, extensions of 
Kolmogorov's entropy proposed by Zurek \cite{zurek89}, 
thermodynamic depth \cite{lloyd88}, Nielsen's circuit 
depth \cite{nielsen10}, and operator growth characterized by
Krylov complexity or state spread complexity \cite{parker19, rabi22, bala22}.
Among these measures, state spread complexity has been introduced 
to be a basis-independent measure; this property arises 
from projecting onto minimal bases via the Lanczos algorithm, 
as part of larger Krylov subspace methods in quantum dynamics 
\cite{rabi25, nandy25pr, baiguera26}.

Quantifying time evolution of quantum complexities, $\mathscr{C}(t)$, 
for black hole dynamics and many-body quantum systems are active 
area \cite{susskind20}. Inspired by Leonard Susskind's 
Princeton lectures \cite{susskind20}, we define a time-integrated quantity, 
$\mathscr{C}_A$ (where $A$ denotes area), 
\begin{equation}
\mathscr{C}_{A} = \int_{0}^{t} \mathscr{C}(\tau) d\tau.
\end{equation} 

Here, we introduce bootstrapping of the operator by adding a 
perturbation: $\mathscr{O}_{A}^{i} =  \mathscr{O}_{A} + \mathscr{O}_{i}$, 
where $\mathscr{O}_{i}$ perturbs the operator in its matrix representation.
This kind of stability analysis for quantum unitary dynamics is pioneered by 
Peres \cite{peres84}.
The resulting set of dynamics generated by the ensemble of perturbed 
operators is suitable for statistical bootstrapping \cite{davison97, efron94}, 
providing physically more robust interval estimates of complexity 
measures and paving the way to practice efficient subsampling \cite{jordans14, bere22}. 
This is not only a statistical procedure but also captures the sensitivity 
of the operator to unitary dynamics. As a result, this new approach allows us to 
generate $M$ integrated complexities: 
$\{\mathscr{C}_{A}^{1}, \mathscr{C}_{A}^{2},...,  \mathscr{C}_{A}^{M}\}$. 
Bootstrapping enables us to estimate statistical properties of the 
generated sample of the complexity measure ensemble. We compute the 
time-integrated fidelities and spread complexities for the quantum state 
evolution of maximally entangled initial states, subject to the operator 
perturbation scheme described above under unitary dynamics $U_i(t)$. 
This procedure generates a set of distinct state evolution paths.
A state based perturbation approach has been applied to thermofield 
double states as the butterfly effect \cite{peres84, shenker14bhbe} and in the 
quantum Lyapunov spectrum \cite{gharibyan19}, that can also generate 
a set of distinct state evolution paths. Our approach relates to multi-seed
Krylov complexity \cite{craps25} without any alteration to Lanczos 
algorithm, focusing on spread complexity instead.

\subsection{Maximally entangled qubits}

We consider an $N$-qubit system prepared in a maximally entangled 
superposition of the global zero and one states \cite{hayden07, 
nielsen10, susskind15qm}:

\begin{equation}
| \psi_N \rangle = \frac{1}{\sqrt{2}} \left( |0\rangle^{\otimes N} + |1\rangle^{\otimes N} \right).
\end{equation}

Explicitly, the tensor product expansion yields:

\begin{equation}
|\psi_N\rangle = \frac{1}{\sqrt{2}} \left( \underbrace{|00\dots0\rangle}_{N} + \underbrace{|11\dots1\rangle}_{N} \right).
\end{equation}

This specific basis state serves as the initial condition for studying 
unitary scrambling with spread complexity, and for subsequent 
fidelity measurements.

\subsection{Unitary quantum dynamics}

The Schrödinger equation for a closed quantum system is given 
by the time-independent eigenvalue problem:
\begin{equation}
H | \psi_{n} \rangle = E_{n} | \psi_{n} \rangle.
\end{equation}
Here, $E_n$ and $|\psi_n\rangle$ denote the eigenenergies and eigenstates 
of the Hamiltonian operator $\hat{H}$. Determining these properties 
for the full Hilbert space constitutes Exact Diagonalization (ED) 
\cite{lin90, nielsen10, susskind15qm}.

Given an initial state $|\psi(0)\rangle$ expanded in this basis as
$$ |\psi(0)\rangle = \sum_{n} c_{n} | \psi_{n} \rangle, \quad \text{with } c_{n} = \langle \psi_{n} | \psi(0) \rangle, $$
the system evolves under the unitary operator $U(t) = e^{-iHt/\hbar}$ 
(setting $\hbar=1$). The state at time $t$ is:
\begin{equation}
|\psi(t)\rangle = U(t)|\psi(0)\rangle.
\end{equation}
In this work, we compute $U(t)$ using spectral decomposition \cite{strang22}:
\begin{equation}
U(t) = V \operatorname{diag}\left(e^{-i E_{n} t}\right) V^{\dagger},
\end{equation}
where $V$ is the matrix of eigenvectors, 
and $\operatorname{diag}(E)$ denotes a diagonal matrix 
with the eigenvalues $E_n$ on its main diagonal.

\subsection{Fidelity of quantum dynamics}

Maintaining quantum correlations \cite{zurek95} is one of the 
core aspects of research in quantum mechanics. 
\cite{sakurai20, nielsen10, susskind15qm}. In our context, 
fidelity $F(t)$ defined to be a measure of the 
similarity between an initial state $|\psi(0)\rangle$ and 
its time-evolved counterpart $|\psi(t)\rangle$. We use 
$F(t)$ as an indicator of information scrambling.

\begin{equation}
F(t) = \left| \langle \psi(0) | \psi(t) \rangle \right|^2.
\end{equation}

$|\psi(0)\rangle$ is the initial configuration, while the 
evolved state is generated by the system's Hamiltonian 
via the unitary operator:
$$ |\psi(t)\rangle = e^{-iHt} |\psi(0)\rangle. $$

The range of $F(t)$ is defined to be  $0 \le F(t) \le 1$. 
The maximum value means theres is no scrambling. At zero, 
it means the final time, the initial state is fully scrambled.

While fidelity functions as a measure of state overlap rather than 
direct complexity or distance, it serves as a fundamental probe 
for scrambling dynamics. 

\section{Analytics for Scrambling} \label{sec:spread}

The concept of spread complexity originates from the study of eigenvalues of linear operators. 
Specifically, it involves a subspace spanned by the repeated action of 
an operator $A$ on an initial state $|\psi_0\rangle$. This subspace 
captures the essential dynamical evolution relevant to the system's 
degrees of freedom. This is the fundamental definition of the Krylov 
subspace \cite{liesen}, efficiently computed via the Lanczos 
algorithm \cite{lanczos50}. Consequently, Krylov spread complexity 
emerges as a robust measure of operator complexity within Hilbert space, 
primarily because it relies on the expansion coefficients in the Krylov 
basis—a metric that is inherently independent of the specific local 
basis chosen for the system \cite{bala22}. While recent 
reviews \cite{baiguera26} have comprehensively covered developments 
in this field, we provide here a concise summary of our specific
implementation analytically. We have introduced the relationship 
between integrated spread complexity and fidelity in probing the 
scrambling of entanglement due to initial condition, using this
Krylov bases formulation of the spread complexity. 

\subsection{Expressions for the Krylov Bases}

The Krylov subspace is defined by the orthonormal basis 
vectors $|K_{n} \rangle$. Here, $n$ is the index of the 
basis vector within the subspace. Its dimension can grow up 
to the Hilbert space dimension of the system acted 
upon by operator $H$. Following the iterative Lanczos
algorithm, at $n=0$, the basis vector $|K_{0} \rangle$ 
corresponds to the initial quantum state $|\psi_0\rangle$ whose 
spread complexity we are investigating. The Lanczos algorithm 
employs two coefficients, $a_{n}$ and $b_{n}$, to determine 
the next basis vector. Consequently, the output of the algorithm 
stores the sequence of vectors $|K_n\rangle$, along with the 
coefficients $b_{n}$ and $a_{n}$.

Initial conditions for $n=0, 1$ must be available before 
we can iterate in computing Krylov bases vectors for $n > 1$. 
We start with $H$ (Hamiltonian matrix) and initial 
state $\psi(0)$. At $n=0$, 

\begin{eqnarray}
b_{0} & = & 0.    \\
|K_{0}\rangle & = & |\psi(0) \rangle, \\       
a_{0} & = &  \langle K_{0}|H| K_{0} \rangle.    
\end{eqnarray}

The next step for $n=1$ is written as follows, 

\begin{eqnarray}
|A_{1} \rangle & = & (H-a_{0} I ) |K_{0} \rangle \\    
b_{1} & = & \langle A_{1} | A_{1} \rangle^{1/2}    \\
|K_{1} \rangle & = & b_{1}^{-1} |A_{1} \rangle   \\
a_{1} & = &  \langle K_{1}|H| K_{1} \rangle  
\end{eqnarray}

Similarly but with a minor tweak, the expressions for $n > 1$ read,

\begin{eqnarray}
|A_{n+1} \rangle & = & (H-a_{n} I ) |K_{n} \rangle - b_{n} |K_{n-1} \rangle \\
b_{n} & = & \langle A_{n} | A_{n} \rangle^{1/2} \\  
|K_{n} \rangle & = &  b_{n}^{-1} |A_{n} \rangle \\
a_{n} & = & \langle K_{n}|H| K_{n} \rangle.   
\end{eqnarray}

The intermediate vectors $A_{n}$ are not stored and 
they only used to update coefficients \cite{lanczos50}. 
The resulting coefficients forms a tri-diagonal matrix
by pairwise $(n, m)$ expectations on the new orthonormal bases
reads, $T_{nm} = \langle K_{n} | H | K_{m} \rangle$.

The basic validation and correctness test would be that 
Krylov vectors $K_{n}$ should form an orthonormal set. 
In the limit of Hilbert space dimension, eigenvalues of 
$T_{nm}$ the tri-diagonal matrix should exactly match those 
of the Hamiltonian $H$. Diagonal and off-diagonal entries 
will be $a_{n}$ and $b_{n}$, respectively. 

We now define the state spread complexity as follows. 
First, we find the Krylov Bases, $|K_{i} \rangle $. 
These vectors can be utilized at a given unitary evolution 
with $|\psi(t) \rangle $. Then the spread complexity defined as 

\begin{equation}
S(t) =   \sum_{n=0}^{m} n | \langle K_{n} | \psi(t) \rangle |^{2}.
\end{equation}

We only need to apply Lanczos algorithm 
once and then evolve the initial state under unitary operator.
$m$ is identified usually at the index where $b_{n}$ value vanishes 
to small number. 

In this work, we propagate the fidelity using exact diagonalization, 
Krylov bases as a matrix, and spectral decomposition for the unitary 
evolution, in computing the spread complexity. However, from 
the basic Schr{\"o}dinger equation perspective in continuous setting,
an iterative differential equation based on Lanczos coefficients 
can also be formed as a one-dimensional chain:  
$$i \partial_{t} \psi_{n}(t) = a_{n} \psi_{n}(t) + b_{n+1} \psi_{n+1}(t) 
+ b_{n} \psi_{n-1}(t).$$
This iterative differential equation is noted
in the literature extensively \cite{parker19, bala22, camargo24}.

\subsection{Integrated Fidelity and Spread Complexity}
\label{sec:krylov-ergo}

We propose to use time-integrated counterparts of fidelity and 
spread complexity in understanding the general relationship between 
fidelity and quantum ergodicity. Usually quantum ergodicity part of the 
relationship involves computing the time correlation of perturb operators 
initially proposed by Prosen \cite{prosen02} in connecting to fidelity
as a foundational contribution. Moreover, recently correlations are used  
in spectral statistics for diagnosing the ergodicity \cite{pathak}. 
By using time-integration at a fixed time horizon $T$, we reduce the effect of 
instantaneous fluctuations, combined with ensemble averaging over bootstrapped 
Hamiltonian realizations, providing a more robust probes.

The time-integrated spread complexity $\mathcal{I}_S(T)$  and 
the integrated infidelity $\mathcal{I}_F(T)$ can be defined
as:
\begin{equation}
\mathcal{I}_F(T) = \int_0^{T} F(t) \, dt, \quad 
\mathcal{I}_S(T) = \int_0^{T} S(t) \, dt.
\end{equation}

The initial maximally entangled state $|K_0\rangle$ is used
in quantifying the survival probability (return amplitude squared), 
$P_0(t) = |\langle K_0 | \psi(t) \rangle|^2$, identical to $F(t)$. 
By using the definition of spread complexity, we can define 
{\it forward spread complexity} $R(t)$, 

\begin{eqnarray}
R(t) = \sum_{n=1}^{m} n | \langle K_{n} | \psi(t) \rangle |^{2} \\
F(t) = S(t) - R(t)
\end{eqnarray}

The time-integrated {\it forward spread complexity} reads:
\begin{equation}
\mathcal{I}_R(T) = \int_0^{T} R(t) \, dt.
\end{equation}

Specific dynamical limits manifest as follows:

\begin{itemize}
    \item \textbf{Ergodic Limit (Chaotic Order):} 
    The scrambling is high so the forward spreading $\mathcal{I}_R(T)^{chaos}$.
    In this regime integrated fidelity drops quickly $\mathcal{I}_F(T)^{chaos}$.
    Hence, $\mathcal{I}_S(T)^{chaos}$ has an inverse trade-off against 
    $\mathcal{I}_F(T)^{chaos}$. 
    \item \textbf{Localized Limit (Integrable Order):} 
    As localization increases, the inverse-trade off continues and saturates, 
     $\mathcal{I}_F(T)^{local}$ reaching to a high-value and forward 
     spreading $\mathcal{I}_R(T)^{local}$ going to zero.
\end{itemize}

We call this inverse trade-off {\it monotonically inverse relationship}
due to the additive nature of spread complexity, being the sum of 
fidelity and forward spreading (scrambling).  

\section{Numerical observations: Rosenzweig-Porter ensemble}\label{sec:rp}

The perturbation scheme developed to generate distinct state 
growth spreading for a given integrable Hamiltonian $H_0$ can 
be expressed as $H = H_0 + H_{\text{per}}$. To validate this framework, 
we utilize the Rosenzweig-Porter (RP) random matrix ensemble \cite{rosen}, 
which has become a standard testbed in many-body studies \cite{kravtsov15}. 
The RP Hamiltonian is constructed by adding two components:

\begin{equation}
H_{\text{rp}} = H_0 + N^{-\gamma/2} H_{\text{goe}},
\end{equation}
 
where $H_{\text{goe}}$ is a random matrix drawn from the Gaussian 
Orthogonal Ensemble (GOE), and $\gamma$ controls the relative weight 
of the off-diagonal disorder. The parameter $\gamma$ serves as a proxy 
for localization strength, interpolating between localized \cite{anderson58} 
and ergodic regimes \cite{lagendijk09}.

In our definition, $H_0$ represents a diagonal matrix with random elements 
drawn from a normal distribution $\text{diag}(H_0) \sim \mathcal{N}(0, 1.0)$, 
while the off-diagonal elements originate solely from the GOE term. 
When $\gamma = 0$, $H_{\text{rp}}$ reduces exactly to a standard GOE matrix, 
corresponding to the fully quantum-chaotic regime with Wigner-Dyson statistics.

Depending on $\gamma$, the ensemble exhibits distinct ergodic 
phases \cite{kravtsov15, pino, jahnke25}: Chaotic (Wigner-Dyson)
$0.0 \le \gamma < 1.0$, Fractal $1.0 < \gamma < 2.0$, and 
Localized (Poisson) $\gamma \ge 2.0$.

This ensemble is prominently used to study many-body localization 
transitions \cite{kravtsov15, pino, tarzia20, tomasi19, khaymovich20,
khaymovich21, sarkar23, bui24}, the analytical structure of eigenstates 
\cite{monthus17, benini, bogo18a, skvortsov22}, and related 
variants \cite{bui26}.

We utilize the RP ensemble as a numerical toy model to investigate integrated 
spread complexity across varying localization strengths defined by $\gamma$. 
Specifically, this setup allows us to probe how an state spreads in 
environments ranging from integrable backgrounds to fully chaotic ones. 
This approach aligns with concepts of the quantum mechanical butterfly 
effect \cite{shenker14bhbe} whereby there state perturbations are used.

\begin{figure}[h!]
\centering
  \includegraphics[width=0.45\textwidth]{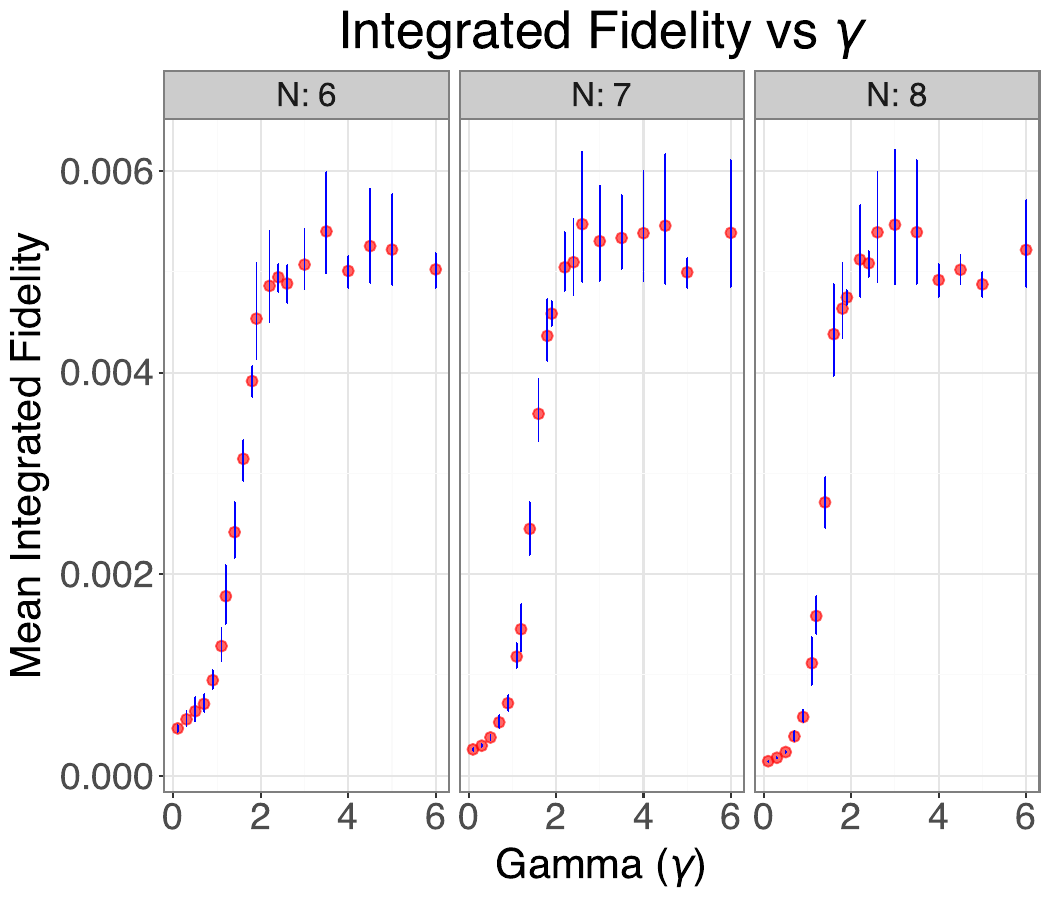}
  \caption{ 
    Integrated fidelity of maximally entangled states evolving under 
    unitary dynamics across the Rosenzweig-Porter ensemble for 
    different order parameters, $\gamma$. Results are obtained using 
    $95\%$ confidence intervals, bootstrapped over 20 
    repeats for each order parameter. We 
    observe the phases here: namely chaotic, fractal and integrable, as 
    a slope, a steeper slope and plateau values for the integrated fidelity. 
  }
  \label{fig-fidelity}
\end{figure}

\subsection{Computing integrated complexity}\label{sec:num}

Numerically exact solutions are produced for unitary time evolution of both 
bootstrapped time-integrated fidelity and spread complexity in tracing the scrambling 
of maximally entangled state, using exact diagonalization with spectral 
decomposition. These are numerically exact because we used full Hamiltonian and 
full Hilbert space for Lanczos iterations. 

Hilbert dimension is defined as $2^{N}$ for a given Hamiltonian. We study 
$N=6,7,8$ for range of different $\gamma$ from $0.1$ to $6.0$ covering 
chaos to integrable regions for the localization strength, or the degree
of complexity. We repeat this $20$ different realizations of the 
Rosenzweig-Porter Hamiltonian per $\gamma$ and computed Krylov bases 
for each realization. Krylov bases are computed with the initial state 
of maximally entangled state.

\begin{figure}[h!]
\centering
  \includegraphics[width=0.45\textwidth]{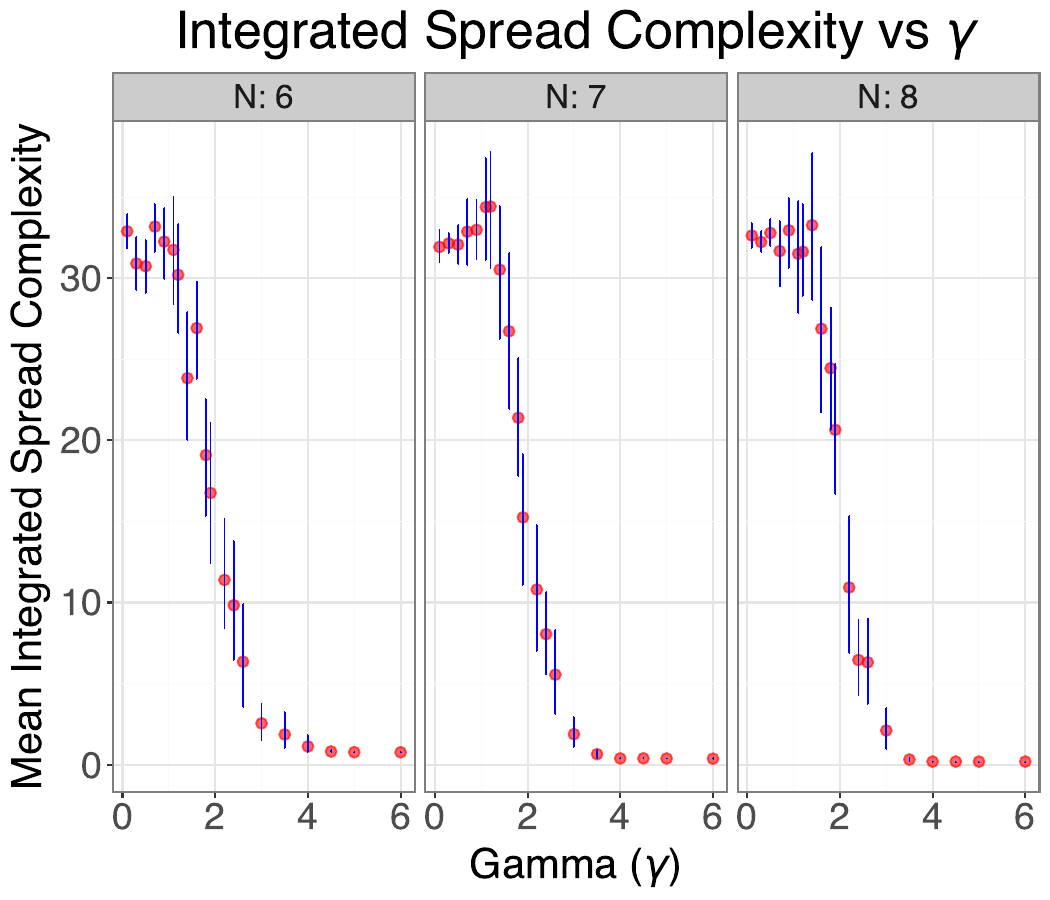}
  \caption{
    Integrated spread complexity of maximally entangled states evolving under 
    unitary dynamics across the Rosenzweig-Porter ensemble for different 
    order parameter, $\gamma$. Results are obtained using $95\%$ confidence 
    intervals; data were bootstrapped over 20 repeats for each 
    order parameter.  We 
    observe the phases here: namely chaotic, fractal and integrable, as 
    plateau, a step downward slop and an other plateau values for 
    the integrated spread complexity. 
  }
  \label{fig-spread}
\end{figure}

\begin{figure}[h!]
\centering
  \includegraphics[width=0.45\textwidth]{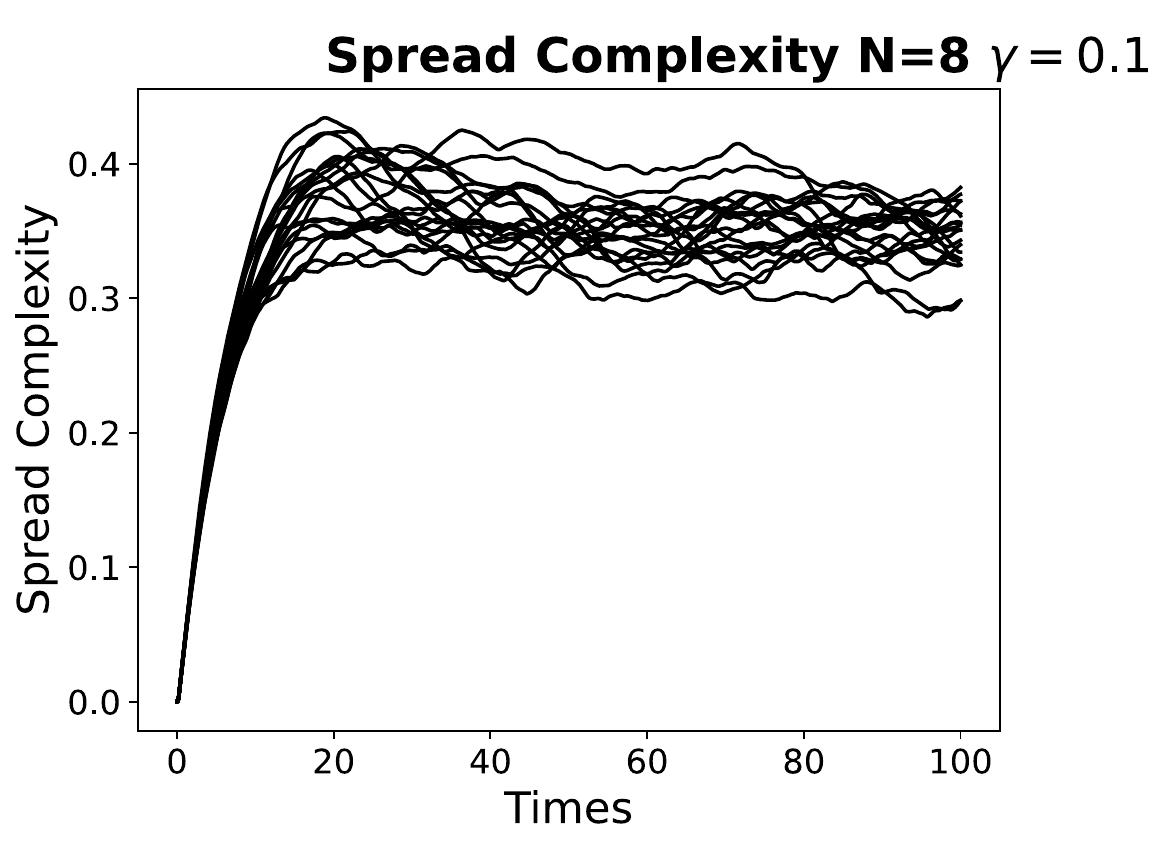}
  \caption{
    Spread complexity of maximally entangled states under 
    unitary evolution over the Rosenzweig-Porter ensemble 
    at $\gamma=0.1$. The inset shows results 
    for 20 different realizations. 
   }
  \label{fig-spread-early}
\end{figure}

The unitary time-evolutions are computed in different scales for 
fidelity and spread complexity, covering their long-times. For fidelity
we use the time step $\delta \tau = 10^{-2}$ with $2000$ steps for the unitary evolution, 
this gives time-integrated fidelities as in Figure \ref{fig-fidelity}. 
Time-integrated fidelity decays rapidly for chaotic cases and gradually 
saturating to a plateau for the full integrable cases. This trend is consistent with 
the physical expectation that in full-integrability we don't see any scrambling but
in thermalized quantum chaos cases scrambling achieved very quickly. 

Integrated spread-complexity shown in Figure \ref{fig-spread}, tracks the 
behavior from chaos to integrability. We use the time step $\delta \tau = 100$ with $50$ 
steps, due to much slower dynamics for this measure compared to fidelity. 
Fast scrambling \cite{susskind08} is observed in highly complex regions and 
then complexity remain small for fully integrable regions.   

\begin{figure}[h!]
\centering
  \includegraphics[width=0.45\textwidth]{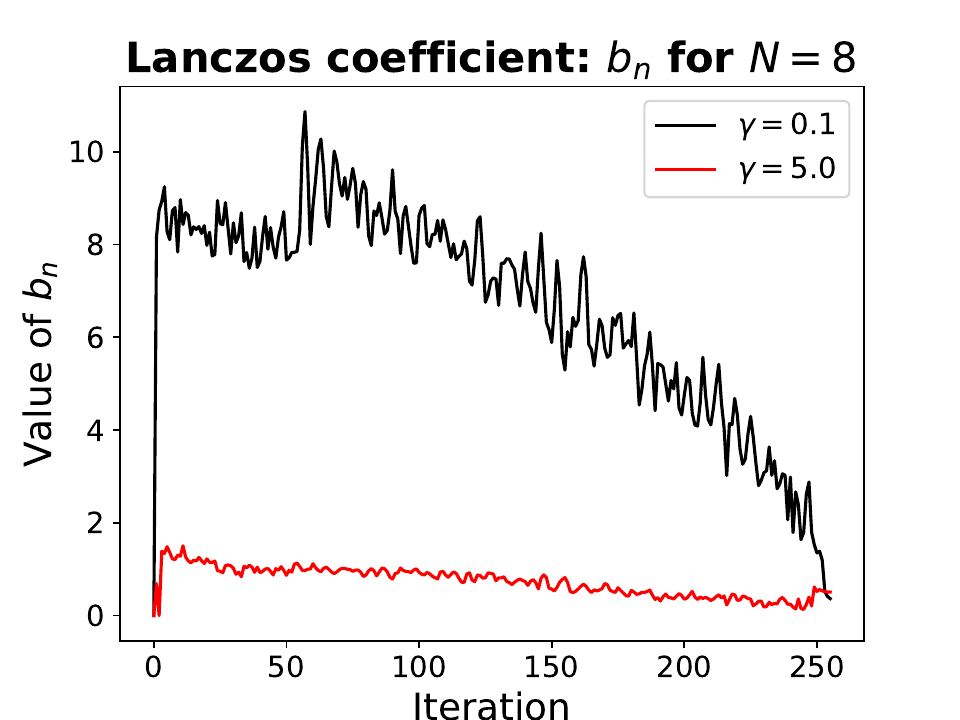}
  \caption{Growth of the Lanczos coefficient $b_{n}$ for chaotic ($\gamma=0.1$) 
  and integrable ($\gamma=5.0$) realizations in the Rosenzweig-Porter 
  ensemble, initialized from superposition states. As both
  operator growth hypothesis \cite{parker19} and state spread \cite{bala22}
  identifies Lanczos $b_{n}$ coefficients would capture the physical 
  characteristics: our observed trends for $b_n$ clearly
  differentiates the chaotic and integrable regimes.}
  \label{fig-lanczos-bn}
\end{figure}

\section{Discussions}\label{sec:dis}

The behavior of spread complexity at fast scrambling at $\gamma=0.1$ for 
$20$ realizations are shown in Figure \ref{fig-spread-early}. As we have discrete 
steps, we simply sum the fidelity and spread complexity values and scale
with $\delta \tau/steps$. Lanczos coefficient $b_{n}$ trends are shown in
Figure \ref{fig-lanczos-bn}, their convergence slope indeed represent 
physical characteristics, chaotic versus regular.

As different time-scales and regimes are extensively studied 
\cite{santos19tr, vidmar20qc}, these regimes exhibit distinct physical 
properties and characteristic relaxation times, namely the Thouless time
for early dynamics and the Heisenberg time for late dynamics. We choose 
step sizes that are small enough to capture early dynamics 
(as in Figure \ref{fig-spread-early}) and a large number of steps 
to reach late times for both observables, calculating integrals 
of fidelity and spread complexity. As the time-window is fixed 
across different ergodicities ($\gamma$ ranges), our results effectively 
represent the global behavior: while they probe the long-time limit, 
due to the integration process, they are also influenced by 
early-time dynamics.

Computing spread complexity with different initial conditions is studied
in-depth, such as thermofield double (TFD) \cite{camargo24, bagg25} and a
particular product states with all zero but the first state is one \cite{bhat25}.
We choose to start with a maximally entangled state in tracking spreading
of maximal entanglement as a clean baseline. Analytical inverse relation we
have shown doesn't depend on the initial state, but preserving a full 
quantum correlation during unitary evolution is the core tenant of the work.
Inverse-relation may diminish when quantum decoherence occurs. Here we follow 
full unitary evolution without decoherence. How this inverse-relationship 
with different initial states of degree of entanglement strengths requires 
more detailed study. 

\section{Conclusions}\label{sec:con}

An extension of spread complexity is introduced and 
studied. Our analytical analysis and numerical studies 
supports the notion that maximally entangled state integrated 
fidelity exhibits monotonically inverse relationship with 
the bootstrapped integrated spread complexity over the 
order parameters. We also shown that integrated quantities 
differentiate all phases of Rosenzweig-Porter ensemble
in this inverse relationship setting.

These physically consistent result signifies a robust 
approach for assessing the degree of quantum ergodicity
in a more accessible and tractable manner in understanding 
fidelity of quantum dynamics to spread complexity and 
spreading of entanglement in a composite view.   \\ \\ 

{\it Acknowledgments:} Y. S\"uzen for her kind support 
in quantum dynamics collaboration and encouragements for 
the development of the Python toolkit Leymosun \cite{suzen25ley}. 
Author is also grateful to Hyun-Sik Jeong, Tanay Pathak, 
Pratik Nandy, Ivan M. Khaymovich and vibrant quantum chaos 
community for their insightful comments and kind correspondence.
Dataset and notebooks are available in the Zenodo 
repository for reproducibility \cite{suzen26sup1}.

\printcredits

\bibliographystyle{unsrtnat}
\bibliographystyle{cas-model2-names}
\bibliography{suzen}





\end{document}